\documentclass[useAMS,usenatbib]{mn2e}

\usepackage{lscape}
\usepackage{graphicx}
\usepackage{txfonts}
\usepackage{multirow}
\usepackage{natbib}
\usepackage{longtable}
\usepackage{colortbl}

\newcommand{\aap}{A\&A}           
\newcommand{\aaps}{A\&AS}         
\newcommand{\aj}{AJ}              
\newcommand{\apj}{ApJ}            
\newcommand{\apjs}{ApJS}          
\newcommand{\mnras}{MNRAS}        
\newcommand{\pasp}{PASP}          

\title[The young stellar cluster DBS2003\,157]{The young stellar cluster [DBS2003]\,157 associated with the
H\,{\sc ii} region GAL\,331.31-00.34\thanks{Based on observations
carried at the SOAR observatory, a joint project of the Minist\'erio
de Ci\^encia, Tecnologia e Inova\c c\~ao (MCTI) of the Rep\'ublica
Federativa do Brasil, the U.S. National Optical Astronomy Observatory
(NOAO), the University of North Carolina at Chapel Hill (UNC), and the
Michigan State University (MSU).}}

\author[M. C. Pinheiro et al.]{
M. C. Pinheiro$^{1,2}$\thanks{E-mail: pinheiro.marcio@gmail.com},
Z. Abraham$^{3}$,
M. V. F. Copetti$^{1}$,
R. Ortiz$^{4}$,
\newauthor
D. A. Falceta-Gon\c{c}alves$^{4}$, and
A. Roman-Lopes$^{5}$ \\
$^{1}$Laborat\'orio de An\'alise Num\'erica e Astrof\'\i sica, Departamento
de Matem\'atica, UFSM, Santa Maria, RS, 97119-900, Brazil \\
$^{2}$Universidade Federal da Fronteira Sul, Campus Cerro Largo, RS, 97900-000, Brazil  \\
$^{3}$Instituto de Astronomia, Geof\'\i sica e Ci\^encias Atmosf\'ericas, Universidade de S\~ao Paulo, Cidade Universit\'aria, S\~ao Paulo, SP, Brazil \\
$^{4}$Escola de Artes Ci\^encias e Humanidades, USP, Av. Arlindo Bettio, 1000,
S\~ao Paulo, SP, 03828-000, Brazil \\
$^{5}$Department of Physics, Universidad de La Serena, Benavente 980, La Serena, Chile}

\begin{document}

\date{Accepted 2012 April 3.  Received 2012 April 2; in original form 2012 January 17}
\pagerange{\pageref{firstpage}--\pageref{lastpage}} \pubyear{2010}

\maketitle

\label{firstpage}

\begin{abstract}
We report a study of the stellar content of the Near-infrared cluster
[DBS2003]\,157 embedded in the extended H\,{\sc ii} region GAL\,331.31-00.34,
which is associated with the IRAS source 16085-5138. $JHK$ photometry was
carried out in order to identify potential ionizing candidates, and the
follow-up NIR spectroscopy allowed the spectral classification of some
sources, including two O-type stars. A combination of NIR photometry
and spectroscopy data was used to obtain the distance of these two
stars, with the method of spectroscopic parallax: IRS\,298 (O6\,{\sc V},
$3.35 \pm 0.61$\,kpc) and IRS\,339 (O9\,{\sc V}, $3.24 \pm 0.56$\,kpc).
Adopting the average distance of $3.29 \pm 0.58$\,kpc and comparing the
Lyman continuum luminosity of these stars with that required to account
for the radio continuum flux of the H\,{\sc ii} region, we conclude that
these two stars are the ionizing sources of GAL\,331.31-00.34. Young stellar
objects (YSOs) were searched by using our NIR photometry and MIR data from
the GLIMPSE survey. The analysis of NIR and MIR colour-colour diagrams
resulted in 47 YSO candidates.

The GLIMPSE counterpart of IRAS\,16085-5138,
which presents IRAS colour indices compatible with an ultra-compact
H\,{\sc ii} region, has been identified. The analysis of its spectral
energy distribution between $2$ and $100\,\mu$m revealed that this source
shows a spectral index $\alpha = 3.6$ between $2$ and $25\,\mu$m, which is
typical of a YSO immersed in a protostellar envelope. Lower limits to the
bolometric luminosity and the mass of the embedded protostar have been
estimated as $L=7.7\times10^3L_{\sun}$ and $M = 10\,M_{\sun}$, respectively,
which corresponds to a B0--B1\,{\sc V} ZAMS star.

\end{abstract}


%
%
\section{Introduction}
\label{Introduction}
The study of the stellar content and the determination of the distances
of H\,{\sc ii} regions and star-forming complexes associated with massive
molecular clouds are fundamental for the determination of the spiral
structure and the rotation curve of the Galaxy. Also, the assessment of
Galactic gradients of chemical abundances and electron temperatures have
strong dependence on the accuracy of the distance estimates. The Norma
region is a very interesting sector of the Galaxy for this investigation,
since the line of sight intersects three spiral arms (Sagittarius-Carina,
Scutum-Crux, and Norma). In addition to the tens of optical H\,{\sc ii}
regions identified by \cite{Rodgers et al 1960}, radio observations
\citep[e.g.][]{Caswell & Haynes 1987} have revealed many other H\,{\sc ii}
regions heavily obscured by interstellar dust. The ionizing stellar
clusters of some of these objects have been studied in detail in the
last decade \citep{Roman-Lopes et al 2003, Roman-Lopes & Abraham 2004,
Russeil et al 2005, Skinner et al 2009, Chavarria et al 2010}. 
However, the ionizing stars of many of the H\,{\sc ii}
regions in the area have not been identified yet, and therefore their
distances remain unknown.

The 2\farcm5 $\times$ 2\farcm5 radio H\,{\sc ii} region GAL\,331.31-00.34
\citep{Caswell & Haynes 1987} is one of the interesting and unexplored
objects located in Norma. It is associated with the IRAS source 16085-5138
and with the $2\farcm1\times1\farcm5$ infrared cluster [DBS2003]\,157
centred at $\rmn{RA}({\rm J2000.0})=16^{\rmn{h}} 12^{\rmn{m}} 20^{\rmn{s}}$,
$\rmn{Dec.}({\rm J2000.0})=-51\degr 46\arcmin 14\arcsec$, which was
identified by \cite{Dutra et al 2003} using data from the 2MASS survey.
Methanol \citep{Ellingsen et al 1996,Kuchar & Clark 1997,Walsh et al 1998,
Caswell et al 2000} and hydroxyl \citep{Caswell et al 1980} maser emission
indicates the existence of a massive star-forming region \citep{Walsh et al
1997}. This is corroborated by the detection of CS(2-1) and SiO molecular
lines \citep[respectively]{Bronfman et al 1996,Harju et al 1998} which
require H$_2$ densities higher than $10^{4}$\,cm$^{-3}$.

GAL\,331.31-00.34 is located at the east part of a large ($1^\circ$)
complex of H\,{\sc ii} regions, which includes eight bright extended
radio sources \citep{Amaral & Abraham 1991}. A high-energy gamma-ray
source, HESS\,J1614-518, was discovered at the edge of this complex
\citep{Aharonian et al 2006}. No counterpart has been found for this
source among the most plausible classes of objects, like pulsar wind
nebulae, supernova remnants, X-ray binaries, or active galactic nuclei.
The superbubbles produced by the strong wind activity of OB associations
could power high-energy gamma-ray sources \citep{Parizot et al 2004}.
Some other extended high-energy sources have also been found associated
with young stellar clusters \citep{Aharonian et al 2007}. To figure out
the energetics of these phenomena, it is essential to obtain the physical
characteristics of the associated stellar population as well as an
accurate estimate of its distance.

In this work, we present $JHK$ photometry of the stellar cluster
[DBS2003]\,157 and near-infrared (NIR) spectrophotometry of nine
selected candidate stellar members. We aim to identify the ionizing
sources of the associated H\,{\sc ii} region GAL\,331.31-00.34 and to
estimate its distance using the spectroscopic parallax method.
In addition to that, $JHK$ photometry and mid-infrared (MIR) data from the GLIMPSE
survey carried out with the Spitzer Space Telescope \citep{Benjamin et al
2003} are used to identify young stellar objects in the area.

\section{Observations and data reduction}
\label{Observations}

\subsection{Imaging photometry}
\label{Photometric data}

$JHK$ photometric observations were performed at the Observat\'orio Pico
dos Dias (OPD), Brazil, in April 2010. We used the NIR Camera CamIV
attached to the 0.6-m Boller \& Chivens telescope to obtain frames
with a field of view of $8\arcmin\times8\arcmin$ and spatial scale
of $0\farcs48$\,pixel$^{-1}$. A Hawaii CCD detector of
$1024\times1024$\,pixels and a set of $J$, $H$, $K_{\rm s}$ filters were
used. The $K_{\rm s}$ narrow band filter {\it C1}
(${\rm FWHM}\sim0.023$\,$\mu$m), centred at 2.138\,$\mu$m, was chosen to
avoid the contamination by Brackett-$\gamma$ nebular emission
\citep[see][]{Roman-Lopes & Abraham 2006a}. To prevent saturation of the brightest stars
in the $J$ and $H$ bands and to minimize the high thermal noise in
the $K_{\rm s}$-band, multiple short exposures were taken in each filter.
 The total integration times were 1260\,s, 1575\,s and 5250\,s in the
$J$-, $H$-, and {\it C1}-bands, respectively. 
In order to subtract the background emission, images were
obtained at five different positions: the centre and its four adjacent positions,
displaced between $90\arcsec$ to $120\arcsec$. These large displacements were
necessary to build a sky image uncontaminated by nebular emission. For each band, the final sky image
represents the median value of each pixel taken over the five dithering
positions.
This technique allows the subtraction of the background  and at the same time preserves the nebular emission in the final images.
The typical seeing ranged
from $1\farcs1$ to $1\farcs5$. Dark exposures and dome flat-fields were
also taken both at the beginning and at the end of each night.

\begin{figure}
\centering
\includegraphics[width=0.36\textwidth,angle=-90]{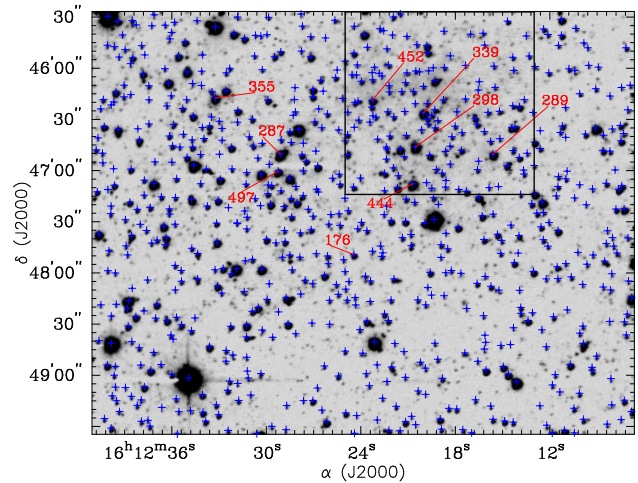}
\caption{CamIV $H$-band image of the cluster [DBS2003]\,157. The $\sim\!2\arcmin\times2\arcmin$ black square surrounds the cluster as indicated in DBS2003. Blue marks denote all the sources found in the field by the {\it daofind} IRAF routine with a detection threshold of $4\sigma$ above the local background. The stars selected for the follow-up spectroscopy are numbered in red.}
\label{fig-cluster157Hdaofind}
\end{figure}

\begin{figure}
\centering\includegraphics[width=0.48\textwidth]{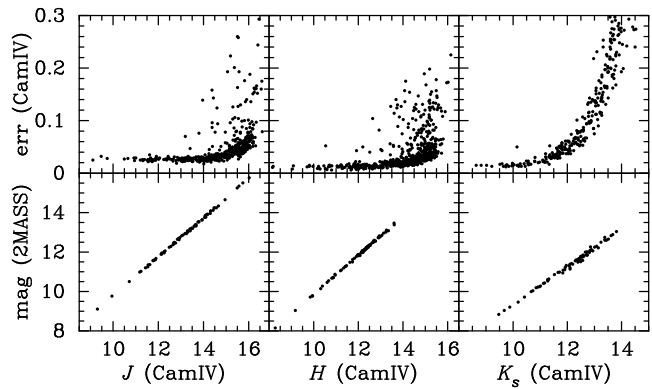}
\caption{Top: Propagated photometric errors versus the $J$, $H$, and $K_{\rm s}$ magnitudes. Bottom: 2MASS versus CamIV magnitudes.}
\label{fig-magXerr}
\end{figure}

Standard procedures for NIR stellar CCD photometry in relatively crowded
fields were performed with the IRAF\footnote {IRAF is distributed by
the National Optical Astronomy Observatory, which is operated by the
Association of Universities for Research in Astronomy, Inc., under
contract to the National Science Foundation.} software. All frames were
individually dark-subtracted and flattened, and the background contribution subtracted.  
Eventually, they  were  aligned and trimmed, to finally generate 
J-, H- and K-band images by median-combining  all frames of each 
filter. By choosing the median-filter, we diminished the influence of 
bad pixels and cosmic rays on the final image.
This procedure resulted in $J$, $H$, and $K_{\rm s}$ ``final'' frames
centred at $\rmn{RA}({\rm J2000})=16^{\rmn{h}} 12^{\rmn{m}} 24^{\rmn{s}}$,
$\rmn{Dec.}({\rm J2000})=-51\degr 47\arcmin 27\arcsec$, covering an area
in the sky of $\sim\!5\arcmin\times4\arcmin$.
Figure~\ref{fig-cluster157Hdaofind} shows one of these frames ($H$-band).
The {\it daofind} IRAF routine was used in the determination of the
physical coordinates of the stars in each frame, adopting an object
detection threshold (in counts) of $4\sigma$ above the local background.

The instrumental magnitudes were computed using the {\it daophot} IRAF
routines. Since some parts of the cluster are too crowded for aperture
photometry, the point spread functions (PSFs) were obtained using the
{\it psf} routine. Simultaneous PSF-fits for all the stars listed by
{\it daofind} were performed with the {\it allstar} routine. A fitting
radius of about $1\times\langle FWHM \rangle$ and a PSF radius of $\sim 3.5$
times this value was assumed for all stars. 

The photometric calibration was performed using data from the 2MASS survey. 
For this task, seventy-three no-blended stars in both 2MASS and CamIV frames 
were chosen outside the $\sim\!2\arcmin\times2\arcmin$ rectangular region 
(see Fig. \ref{fig-cluster157Hdaofind}) centred at the coordinates of 
[DBS2003]\,157. Using the least squares method, we adjusted linear 
transformation equations for converting the instrumental magnitudes to 
the 2MASS system. The standard deviations of these transformations, defined as
\begin{equation}
\sigma^2 = \frac{1}{n-2}\,\sum_{i}{ \left[m_{\rm 2MASS}^{(i)} - \beta_0 - \beta_1\,m_{\rm CamIV}^{(i)} \right]^2}\,,
\end{equation}
were 0.034, 0.026, and 0.036 for the $J$, $H$, and $K_{\rm s}$ (C1) filters, 
respectively. The final photometric errors were obtained as the quadratic 
sum of the transformation errors and of the individual error in the 
instrumental magnitudes given by the {\it allstar} task. No significant 
$(J-H)$ and $(H-K)$ colour-dependence between 2MASS and CamIV photometric 
systems has been found \citep{Roman-Lopes & Abraham 2006b}. Figure~\ref{fig-magXerr} 
shows the linear relation observed between the 2MASS and CamIV magnitudes and the
dependence of the errors on the photometric magnitude for all stars detected.
By analysing the magnitude histograms, $\log(N)$ vs. $m$, we estimate the
completeness limits of the photometry as $J=15.5$\,mag, $H=15.2$\,mag,
and $K_{\rm s}=13.5$\,mag, defined as the point where the $\log(N)$ vs. $m$ plot
deviates from a straight line. These values are similar to the 2MASS limits
on the $J$ and $H$ bands ($15.8$ and $15.1$\,mag, respectively), but the $K_{\rm s}$ 
completeness limit of the CamIV data is $0.8$\,mag lower.

Table \ref{tab-photometry} presents the
results of the photometry. Columns 1 lists our identification of the
star; columns 2 and 3 provide the equatorial coordinates obtained with
the {\it ccmap} IRAF routine applied to pixel-coordinates listed by
the {\it daofind}; columns 4, 5 and 6 list the $J$, $H$, and $K_{\rm s}$
photometric magnitudes and errors.

At a spatial resolution twice as better as the 2MASS survey, we
to identified many sources that were previously unresolved in the 2MASS
data, which in turn improves significantly the photometric accuracy,
especially for stars situated in the crowded areas.

\subsection{Spectroscopy}
\label{spec-data}
NIR spectroscopic data of nine stars in the GAL\,331.31-00.34 region
were acquired with the Ohio State Infrared Imager/Spectrometer
(OSIRIS\footnote{OSIRIS is a collaborative project between the Ohio State
University and Cerro Tololo Inter-American Observatory (CTIO).}) coupled
to the 4.1-m telescope of the Southern Observatory for Astrophysical
Research (SOAR), located at Cerro Pachon, Chilean Andes. Spectra in
$J$-, $H$-, and $K_{\rm s}$-band were taken with OSIRIS in the low-resolution
(R $\simeq$ 1200) multi-order cross-dispersed (XD) mode. In this mode,
the instrument operates with a f/2.8 camera and a short (27\arcsec $\times$
1\arcsec) slit, covering the three bands simultaneously, in adjacent orders.
The raw spectra were acquired using the standard AB nodding technique.
Multiple short exposures were taken at each nod position, totalling 8-to-20
individual frames, giving a ``final'' exposure time of 8-to-40 minutes.
Table~\ref{List of spectroscopic targets} presents a journal of the
spectroscopic observations. It lists our identification of the star,
as defined in Table~\ref{tab-photometry}, the identification in the
2MASS catalogue, the number and the duration of the individual exposures,
and the observation dates. All the target stars were OB candidates selected
from our photometry, with one exception: IRS\,339, observed in 2008 and
selected from an initial analysis based on the 2MASS data.

\begin{table}
\caption{Journal of spectroscopic observations}
\begin{tabular}{@{}lllllllllll}
\hline\hline
\multicolumn{2}{c}{ID}      & Indiv. exp.      & N$^{\rm o}$ of & \multicolumn{1}{c}{Date}     \\
\cline{1-2}
\noalign{\smallskip}
  This work &\multicolumn{1}{c}{2MASS} & time (s)  &  exp.  &    \\
\hline
\noalign{\smallskip}
IRS\,339  & J16122002-5146262    & 120  & 16  &  2008/07/10   \\
IRS\,298  & J16122053-5146460    & 75   & 12  &  2011/05/08   \\
IRS\,287  & J16122911-5146503    & 75   & 12  &  2011/05/08   \\
IRS\,497  & J16122925-5147004    & 120  & 12  &  2011/06/02   \\
IRS\,355  & J16123324-5146173    & 100  & 08  &  2011/06/02   \\
IRS\,176  & J16122445-5147492    & 120  & 12  &  2011/06/03   \\
IRS\,444  & J16122071-5147075    & 90   & 08  &  2011/06/03   \\
IRS\,289  & J16121561-5146504    & 75   & 20  &  2011/06/12   \\
IRS\,452  & J16122326-5146188    & 120  & 20  &  2011/06/12   \\
\hline
\label{List of spectroscopic targets}
\end{tabular}
\end{table}

A- and G-type spectroscopic standard stars were observed immediately
before and after the ``science'' targets at similar air masses in order
to remove telluric atmospheric absorption effects from the ``science''
spectra. These raw spectra were reduced using the CIRRED package and
usual IRAF tasks. Two-dimensional frames were sky-subtracted for each
pair of images taken at the two nod positions A and B, followed by
dividing of the resultant image by a master flat. Thereafter, wavelength
calibration was applied using sky lines; the typical error (1\,$\sigma$)
for this calibration is estimate as $\sim$12\,\AA. The multiple exposures
were combined and one-dimensional spectra were generated.

Telluric atmospheric correction using the spectroscopic standard stars
completed the reduction process. In this last step, we divided each
``science'' spectrum by the spectrum of the A0\,{\sc V} spectroscopic
standard star free of photospheric features, carefully removed by
interpolating across their wings using the continuum points on both sides
of the line. In the case of the $H$-band, the subtraction of the
hydrogen absorption lines could not be successfully done by this
method because of the small separation between the multiple lines of the
Brackett series and some strong telluric features. Therefore we proceeded
to remove the hydrogen lines from the A0\,{\sc V} $H$-band spectrum
using the technique used by \cite{Blum et al 1997}. Basically, one obtains 
a spectrum composed of only the profiles of hydrogen lines dividing the 
spectrum of a A-type standard star by a spectrum of a G-type star, whose H\,{\sc i} 
intrinsic lines (fairly weaker than in A0\,{\sc V} line spectra) and other features 
of G-type spectra were previously 
removed by hand, using as template the NOAO solar atlas of \cite{Livingston & Wallace 1991}. 
After that, these profiles are used to correct the A0\,{\sc V} spectrum, generating 
an $H$-band spectrum free of the hydrogen Brackett lines. Finally, telluric bands were
removed using the IRAF task {\it telluric}. This algorithm interactively
minimizes the RMS in specific regions of the sample by adjusting the
wavelength shifts and intensity scales between the standard and ``science''
spectrum before performing the division. The wavelenght shifting corrects
possible errors in the dispersion zero-points, whereas the intensity
scaling equalizes airmass differences and variations in the abundance
of the telluric species. Typical values of the shifts varied around
$\sim$2\,\AA, while the scaling factors were smaller than 10\%.

\begin{figure*}
\centering\includegraphics[height=16cm,angle=-90]{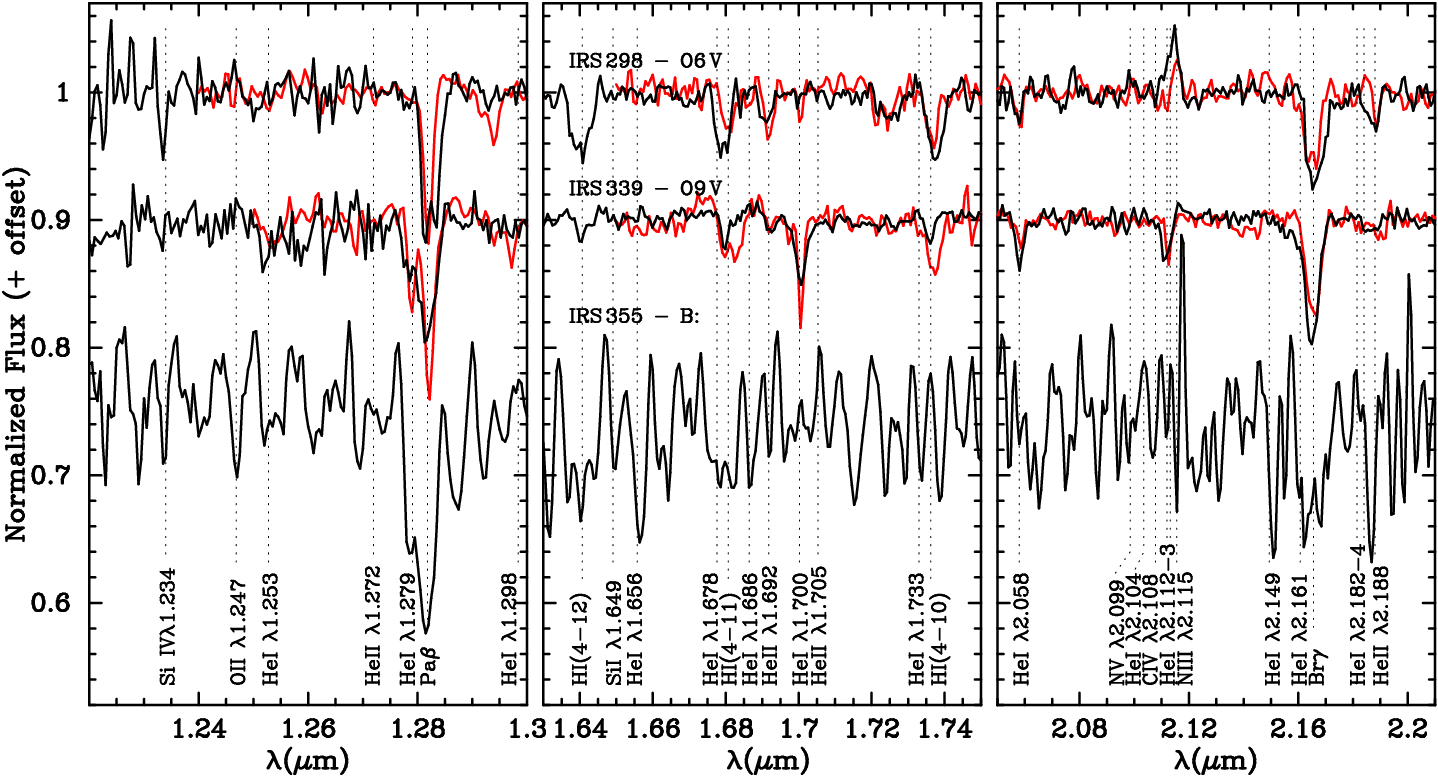}
\caption{Spectra of the hot stars identified in the field of GAL\,331.31-00.34
(in black) and the comparison spectra for a O6\,{\sc V} star (HD\,101190) and
a O9\,{\sc V} star (HD\,93028) (in red).}
\label{fig-spec-classif}
\end{figure*}

\section{Results}
\label{Results}

\subsection{Spectral classification}
\label{Spectral classification}

{He\,{\sc i}} optical absorption lines are well-known spectral features of
OB stars \citep[cf.][]{Walborn & Fitzpatrick 1990}. If He\,{\sc ii}
absorption lines are also present, there is a substantial indicative
of an O-type star. Since such features are also found in NIR spectra
\citep{Hanson et al 1996}, the presence of these lines must still be
the primary criterion to be observed when classifying hot star spectra
in this band. H\,{\sc i} Brackett series could also be used. However, 
it would require special attention since these lines are highly contaminated 
by nebular emission \citep{Bik et al 2005}.

\begin{figure}
\includegraphics[width=59mm,angle=-90]{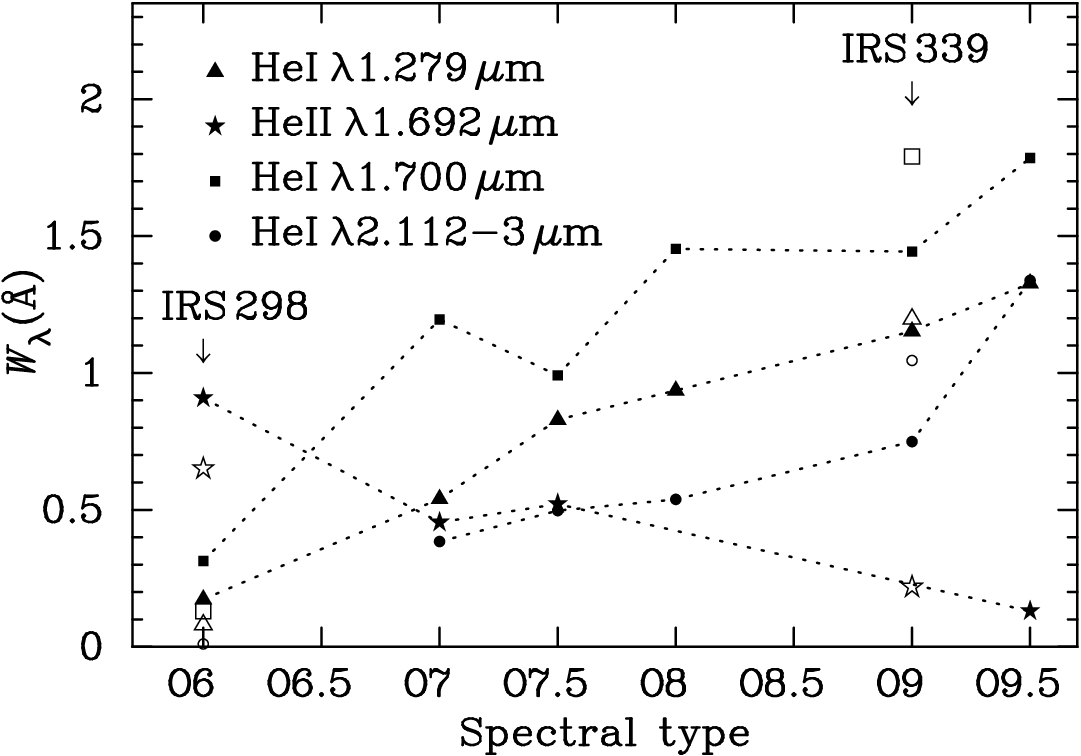}
\caption{Equivalent width $W_\lambda$ of He-lines versus the spectral type of dwarf stars (filled symbols) and a comparison with the values of $W_\lambda$ measured for IRS\,298 and IRS\,339 (open symbols, plotted at the spectral types O6 and O9, respectively).}
\label{ew-classif}
\end{figure}

The spectra of six of the stars listed in Table
\ref{List of spectroscopic targets} were found to correspond to
late-type objects. They were classified visually by comparison with
the 0.8-5 $\micron$  spectral library of cool stars by \cite{Rayner et al 2009}
and are displayed in Fig.~\ref{fig-spec-cold-stars} (Appendix). They are
probably field stars and will not be included in our discussion on the
region. The other three spectra correspond to early-type stars
(Figure~\ref{fig-spec-classif}).

The detection of He\,{\sc ii} lines at 1.692\,$\mu$m and 2.188\,$\mu$m
in the spectra of IRS\,298 and IRS\,339 restricts the spectral types of
both of these stars to O9 or earlier. On the other hand, the relative
strength of these lines compared with the He\,{\sc i} features at
1.279\,$\mu$m, 1.700\,$\mu$m and 2.112\,$\mu$m indicates that IRS\,339
is a late-O, whereas IRS\,298 is an early-O star.
The detection of the N\,{\sc iii} line at 2.115\,$\mu$m in emission,
together with the broad profile of the H\,{\sc i} lines, indicates that
these two objects are dwarf stars, even though we do not discard that
IRS\,339 may be a supergiant. The distinction between the various
luminosity classes can be doubtful in the NIR because of the paucity of
lines suitable for this purpose and the large uncertainties of their
equivalent widths \citep{Hanson et al 1996,Hanson et al 2005}.

Figure~\ref{ew-classif} presents a comparison between the equivalent
widths ($W_{\lambda}$) of four selected He lines of dwarf stars, ranging
from O6 to O9.5 spectral types, along with the
values measured for IRS\,339 and IRS\,298. In this comparison, we
included only lines with equivalent widths reasonably well determined
and discarded H\,{\sc i} lines, whose profiles are highly dependent
on the reduction process (see Sec. \ref{spec-data}). Based on this
scheme, the arguments presented above, and a visual comparison with
other dwarf O-star spectra obtained with the same instrumental
configuration and the spectra of the atlases by
\cite{Hanson et al 1996,Hanson et al 2005} and \cite{Bik et al 2005},
we classified IRS\,298 as O6\,{\sc V} and IRS\,339 as O9\,{\sc V}.

The third stellar spectrum displayed in Fig.~\ref{fig-spec-classif}
is too noisy to be accurately classified. However, the strength of
its He\,{\sc i} lines and the tentative detection of Si, C, N, O
suggest that IRS\,335 is an O9-to-B1 type star.

\subsection{Spectroscopic parallax distance}
\label{parallax distance}

The distances of the classified stars IRS\,298 and IRS\,339 have been
calculated from their apparent $J$ and $H$ magnitudes listed in Table
\ref{tab-photometry} using the method of spectroscopic parallax.
The IR extinction curve by \cite{Stead & Hoare 2009}, which results in
$E_{J-H}/E_{H-K} = 1.91$, was adopted for the reddening corrections.
As intrinsic stellar parameters, we used the absolute magnitudes for
the various spectral types compilated by \cite{Vacca et al 1996} and
the non-reddened colour indices by \cite{Koornneef 1983}, both
transformed into the 2MASS photometric system using the relations given
by \cite{Carpenter 2001}. According to this method, we obtained a
heliocentric distance of $d = 3.24\pm0.56$\,kpc ($A_J=3.03, A_V=10.44$)
for IRS\,339 and $d = 3.35\pm0.61$\,kpc for IRS\,298 ($A_J=3.51, A_V=12.10$).
Based on the average distance of these two stars, we estimate the
heliocentric distance for the H\,{\sc ii} region GAL\,331.31-00.34 of
$d=3.29\pm0.58$\,kpc and a visual extinction of
$\langle A_V \rangle \sim 11$\,mag.

\citet{Caswell & Haynes 1987} obtained the radial velocity of
$V_\mathrm{LSR} = -64$\,km\,s$^{-1}$ for GAL\,331.31-00.34, based on
radio recombination lines. According to the model of
\cite{Brand & Blitz 1993} for the rotation curve of the Galaxy, this
figure corresponds to the near and far kinematic distances of $4.2$\,kpc
and $10.7$\,kpc, respectively. Thus this H{\sc ii} region is definitely
at the near kinematic distance.

\begin{figure}
\includegraphics[width=0.46\textwidth,angle=0]{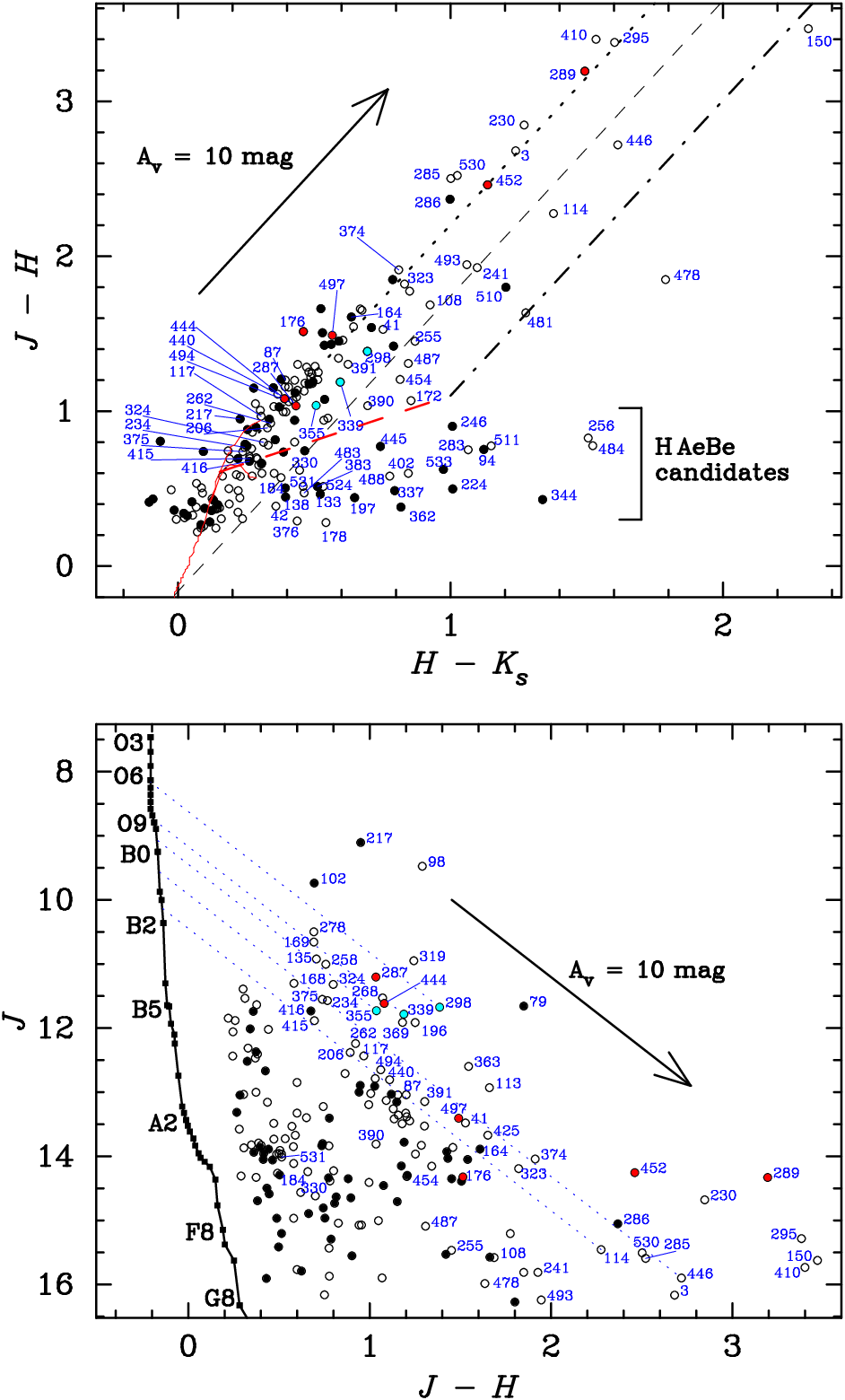}
\caption{Top: $JHK$ colour-colour diagram. The solid red line depicts the main and the giant sequences (only from G3{\sc III}-type on), the dashed red line represents a classical T-Tauri star, and the dotted, dashed, and dot-dashed black lines show the reddening lines for M-type, O-type dwarf stars, and classical T-Tauri, respectively (References in section \ref{ionizing stars}). Bottom: colour-magnitude diagram. The black line represents the main sequence for the distance of 3.29\,kpc. In both diagrams, the filled circles refer to the stars in the black square shown in Fig. \ref{fig-cluster157Hdaofind}, whereas the open circles refer to the other stars on the frame. The spectroscopically classified stars are indicated by blue (early-type) or red circles (late-type). Reddening lines for the 5 potential ionizing stars are also indicated.}
  \label{CC and CM diagrams}
\end{figure}

\subsection{Inventory of the ionizing stars and the Lyman continuum luminosity}
\label{ionizing stars}

In order to search other potential ionizing stars besides those confirmed
spectroscopically, we reanalysed the NIR colour-colour (CC) and
colour-magnitude (CM) diagrams (Figs. \ref{CC and CM diagrams}),
assuming the distance of 3.29\,kpc obtained in section
\ref{parallax distance}. The intrinsic colours assumed are given by
\cite{Koornneef 1983}, while the absolute magnitudes in $J$-band are
from \cite{Vacca et al 1996} for O-stars and from \cite{Wegner 2007}
for the other spectral types, all of them transformed into the 2MASS
photometric system by the relations given by \cite{Carpenter 2001}.
The classical T-Tauri star locus shown in the CC diagram is taken from
\cite{Meyer et al 1997}, while the locus of the Herbig Ae/Be candidates 
is indicated in accordance with \cite{Lada & Adams 1992}. 
According to the CC diagram, we suspect that
IRS\,114 and IRS\,446 are stars with spectral types between B0\,{\sc V}
and B1\,{\sc V}. However, they are heavily obscured by dust 
($A_V > 18$\,mag) and, in this region of the diagram, the classification is
highly dependent on the slope adopted for the reddening vector.
\cite{Nishiyama et al 2006} and \cite{Stead & Hoare 2009} showed that 
this slope depends on the direction of observation and, for large 
reddening, the scattering around the reddening vector might be 
very high. In these cases, a purely-photometric classification 
may not be very accurate. Thus, it could not be discarded  
that IRS\,114 and IRS\,446 actually are pre-main sequence stars.

An inventory of our search for hot stars in the field is presented in
Table 2, which lists the five stars identified as
potential ionizing sources of GAL\,331.31-00.34 along with their
photometric and spectroscopic spectral classifications.

\begin{table}
\caption{Ionizing stars identified in GAL\,331.31-00.34}
\begin{tabular}{@{}lllllllllll}
\hline\hline
\multicolumn{2}{c}{Star ID}          &     & \multicolumn{2}{c}{Classifications}\\
\cline{1-2}\cline{4-5}
\noalign{\smallskip}
This work &\multicolumn{1}{c}{2MASS}& & Photometrical & Spectral           \\
\hline
\noalign{\smallskip}
IRS\,298  & J16122053-5146460    &    &  mid-O        &  O6\,V             \\
IRS\,339  & J16122002-5146262    &    &  late-O       &  O9\,V             \\
IRS\,355  & J16123324-5146173    &    &  B0:          &  B0:               \\
IRS\,114  & J16123539-5148304    &    &  B1:          & $\cdots$           \\
IRS\,446  & J16120863-5146482    &    &  B0:          & $\cdots$           \\
\hline
\label{ionizing sources}
\end{tabular}
\end{table}

To test the completeness of the set of ionizing stars found associated with
this H{\sc ii} region, we compare the Lyman continuum photon luminosity
$N_{\rm Ly}$ (photon s$^{-1}$) emitted by these stars with that needed to
supply the radio continuum flux. Adopting the Lyman luminosity by
\cite{Hanson et al 1997}, these five stars would amount to a total of
$N_{\rm Ly} = 10^{49.07}$\,photons\,s$^{-1}$. In fact, this total is
strongly dominated by the contribution of the two O-type stars. On the
other hand, a Lyman continuum luminosity of $N_{\rm Ly} = 10^{48.74}$
photons\,s$^{-1}$ is inferred from the radio continuum flux density
$S_{\!\nu}$ at $22$\,GHz measured by \cite{Amaral & Abraham 1991},
using the expression: \citep[see][]{Rubin 1968}
\begin{equation}
N_{\rm Ly} = \frac{5.59\times10^{48}}{1+f_i\langle {\rm He^+ / (H^+ + He^+)} \rangle}\left( \frac{\nu}{5\,{\rm GHz}} \right)^{0.1} T_\mathrm{e}^{-0.45} S_{\!\nu} d^2,
\end{equation}
where we assume the abundance ratio of ${\rm He}^+/{\rm H}^+=0.5$, a
typical electron temperature $T_\mathrm{e} = 8000$\,K, and $f_i = 0.65$
for the fraction of He-recombination photons energetic enough to ionize
the H. Using the flux densities at 5\,GHz measured by \cite{Caswell & Haynes 1987} and \cite{Kuchar & Clark 1997} almost identical results are found. Thus, we conclude that the O-type stars IRS\,298 and IRS\,339 can cope with the ionization of this nebula. The three stars photometrically
classified as B0-to-B1\,V type and other cooler stars would not contribute
significantly to the ionization of the nebula.

\subsection{YSO candidates}
\label{YSO candidate}

The intrinsic IR excess found in some stellar objects is an evidence
of accreting disks or infalling envelopes. The identification of a group
of objects with these properties indicates the existence of intense and
recent star formation, which makes these regions interesting places to
search for pre-main-sequence stellar objects (PMS), for example. The
analysis of NIR and MIR colour-colour diagrams allows us to identify
PMS stars and infer their nature based on their photometric properties.
For example, according to \cite{Allen et al 2004}, objects with
accreting disks, named Class II YSO, occupy the locus of reddened
classical T-Tauri in NIR colour-colour diagrams, whereas objects whose
emission is dominated by infalling envelopes, named Class I, have more
reddened colours. YSOs with weak or no IR-excess are named Class III,
and are not discriminated in CC diagrams, since they are found along
with ZAMS stars.

\begin{figure}
\centering
\includegraphics[width=0.36\textwidth,angle=-90]{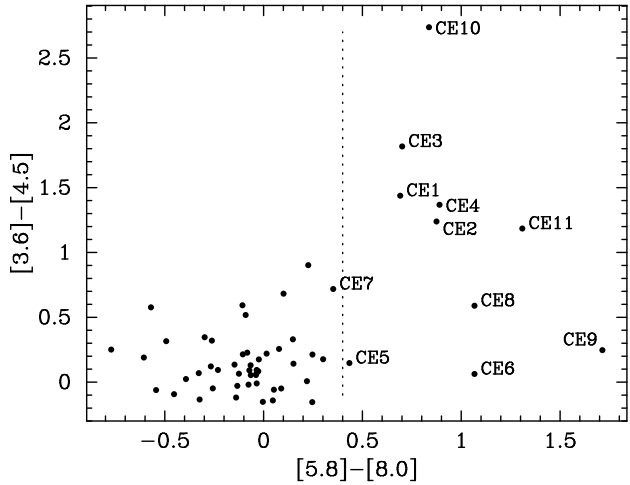}
\caption{MIR colour-colour diagram of stars in a region with $2\farcm5$ of radius and centre matching with $H$-band frame presented in Fig. \ref{fig-cluster157Hdaofind}. YSOs with MIR colour excesses lie rightward of the dotted line. Class III YSOs and back/foreground objects must fall in the clump around the $(0,0)$ point.}
  \label{CC IRAC diagram}
\end{figure}
\begin{table}
\centering
\caption{Young Stellar Objects candidates from the NIR photometry}
\begin{tabular}{@{}lcclcccccc}
\hline\hline
\multicolumn{3}{c}{Identifier}                        &  Class.    \\
\cline{1-3}
\noalign{\smallskip}
 This work &       2MASS        &       GLIMPSE       &            \\
\hline
\noalign{\smallskip}
 IRS\,42   & J16123385-5149269  & G331.3195-00.3979   & H\,AeBe    \\
 IRS\,94   & J16122500-5148459  &     $\cdots$        & H\,AeBe    \\
 IRS\,114  & J16123539-5148304  & G331.3330-00.3892   &  T\,Tau$^\dagger$     \\
 IRS\,133  & J16121385-5148204  & G331.2944-00.3491   & H\,AeBe    \\
 IRS\,138  & J16122037-5148157  & G331.3076-00.3597   & H\,AeBe    \\
 IRS\,150  & J16123309-5148076  & G331.3331-00.3806   & T\,Tau     \\
 IRS\,172  & J16120964-5147550  &    $\cdots$         & T\,Tau     \\
 IRS\,178  & J16123437-5147487  & G331.3391-00.3790   & H\,AeBe    \\
 IRS\,184  & J16121891-5147434  & G331.3110-00.3506   & H\,AeBe    \\
 IRS\,197  & J16121395-5147382  & G331.3025-00.3409   & H\,AeBe    \\
 IRS\,224  & J16121561-5147255  &     $\cdots$        & H\,AeBe    \\
 IRS\,246  & J16121386-5147139  & G331.3071-00.3357   & H\,AeBe    \\
 IRS\,256  & J16122966-5147088  &     $\cdots$        & H\,AeBe    \\
 IRS\,283  & J16123053-5146525  &     $\cdots$        & H\,AeBe    \\
 IRS\,330  & J16122692-5146314  &     $\cdots$        & H\,AeBe    \\
 IRS\,337  &     $\cdots$       &     $\cdots$        & H\,AeBe    \\
 IRS\,344  & J16122121-5146255  &     $\cdots$        & H\,AeBe    \\
 IRS\,362  & J16122064-5146136  &     $\cdots$        & H\,AeBe    \\
 IRS\,376  & J16121207-5146026  &     $\cdots$        & H\,AeBe    \\
 IRS\,383  & J16121868-5145597  &     $\cdots$        & H\,AeBe    \\
 IRS\,390  & J16123848-5145538  & G331.3687-00.3630   & T\,Tau     \\
 IRS\,402  & J16123964-5145474  &     $\cdots$        & H\,AeBe    \\
 IRS\,445  & J16122483-5146507  &     $\cdots$        & H\,AeBe    \\
 IRS\,446  & J16120863-5146482  & G331.3022-00.3213   &  T\,Tau$^\dagger$     \\
 IRS\,454  & J16122704-5146168  & G331.3427-00.3474   & T\,Tau     \\
 IRS\,478  & J16122996-5148154  & G331.3258-00.3766   & H\,AeBe    \\
 IRS\,481  & J16122752-5148042  & G331.3232-00.3700   & T\,Tau     \\
 IRS\,483  & J16121153-5147565  & G331.2947-00.3402   & H\,AeBe    \\
 IRS\,484  & J16122607-5147546  & G331.3223-00.3655   & H\,AeBe    \\
 IRS\,487  & J16122611-5147500  & G331.3232-00.3646   & T\,Tau     \\
 IRS\,488  & J16120772-5147483  & G331.2891-00.3319   & H\,AeBe    \\
 IRS\,510  & J16122112-5146395  & G331.3274-00.3416   & T\,Tau     \\
 IRS\,511  & J16120732-5146382  &     $\cdots$        & H\,AeBe    \\
 IRS\,524  & J16123883-5148207  & G331.3413-00.3932   & H\,AeBe    \\
 IRS\,531  & J16120752-5146336  & G331.3028-00.3164   & H\,AeBe    \\
 IRS\,533  &     $\cdots$       &     $\cdots$        & H\,AeBe    \\
\hline
\multicolumn{4}{l}{ $\dagger$See Section \ref{ionizing stars}.}
\label{NIR YSOs}
\end{tabular}
\end{table}

\begin{table}
\centering
\caption{Identifiers of the YSOs candidates from the MIR GLIMPSE photometry}
\begin{tabular}{@{}lcccccccc}
\hline\hline
\multicolumn{2}{c}{This work}& & \multicolumn{2}{c}{Surveys}     \\
\cline{1-2} \cline{4-5}
\noalign{\smallskip}
 MIR    &       NIR    & &       GLIMPSE        &       2MASS              \\
\hline
\noalign{\smallskip}
\noalign{\smallskip}
  CE\,1    & $\cdots$     & & G331.2999-00.3784    &  J16122314-5149240   \\
  CE\,2    & $\cdots$     & & G331.3093-00.3762    &  J16122521-5148551   \\
  CE\,3    & $\cdots$     & & G331.3205-00.3817    &     $\cdots$         \\
  CE\,4    & IRS\,552     & & G331.3131-00.3653    &  J16122341-5148167   \\
  CE\,5    & IRS\,168     & & G331.3298-00.3735    &  J16123029-5147575   \\
  CE\,6    & IRS\,176     & & G331.3203-00.3615    &  J16122445-5147492   \\
  CE\,7    & $\cdots$     & & G331.3341-00.3619    &  J16122844-5147163   \\
  CE\,8    & IRS\,645     & & G331.3541-00.3664    &  J16123527-5146387   \\
  CE\,9    & IRS\,355     & & G331.3543-00.3585    &  J16123324-5146173   \\
 CE\,10    & $\cdots$     & & G331.3416-00.3464    &  J16122648-5146177   \\
 CE\,11    & $\cdots$     & & G331.3497-00.3497    &     $\cdots$         \\
\hline
\label{MidIR YSOs}
\end{tabular}
\end{table}
\begin{figure}
\centering
\includegraphics[width=0.32\textwidth,angle=-90]{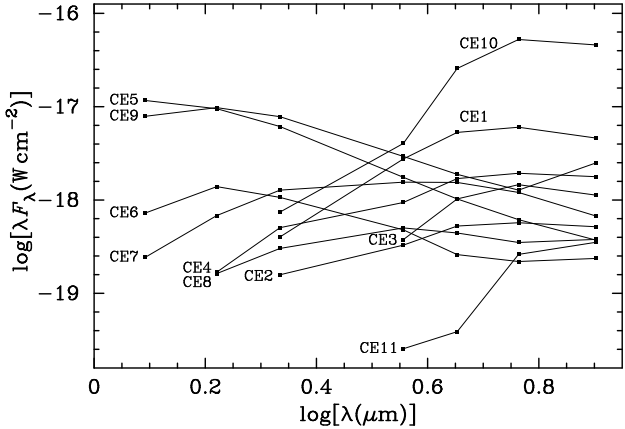}
\caption{Spectral energy distributions (between 1.25 and 10\,$\mu$m) of objects with MIR colour excess. Photometric data extracted from the 2MASS and GLIMPSE surveys.}
  \label{YSO SEDs}
\end{figure}
\begin{figure}
\includegraphics[width=0.32\textwidth,angle=-90]{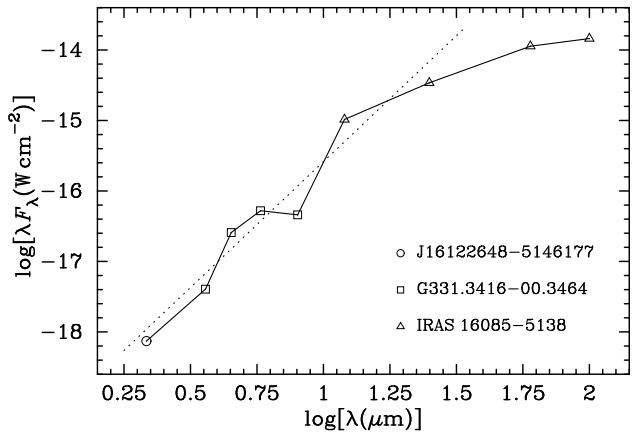}
\caption{Spectral energy distribution (between 1.25 and 100\,$\mu$m) of the source CE\,10 based on data extracted from the 2MASS, GLIMPSE, and IRAS surveys. The dotted line (with spectral index of $\alpha \approx 3.6$) represents the fitting of the SED between $2$ and $25\,\mu$m.}
  \label{IRAS SED}
\end{figure}

The analysis of the NIR colour-colour diagram presented in
Fig.~\ref{CC and CM diagrams} indicates a total of 36 YSO candidates,
which are listed in Table \ref{NIR YSOs} together with their tentative
classification between T-Tauri (T\,Tau) or Herbig Ae/Be (H\,AeBe) stars.
In this list, IRS\,446 and IRS\,114 could also be main sequence stars
very obscured by dust. IRS\,150 is the most reddened object detected
in this work, with $A_V \ga 20$\,mag. Considering the error bars of the
photometry and extinction laws, IRS\,150 has colours of a very reddened
T-Tauri star. Redwards in the CC NIR diagram, IRS\,478 presents typical
colours of a very embedded YSO in an early stage of formation.
A third group of stars appears below the T-Tauri locus, showing
$(H-K_s) \ga 0.4$\,mag. In this region, Herbig Ae/Be stars, which are
PMS stars more massive than T-Tauri, can be confused with Class I sources 
\citep{Lada & Adams 1992}, galaxies or other background objects, making a 
classification based only on photometric data unreliable.

Figure~\ref{CC IRAC diagram} presents a MIR colour-colour diagram
made with data from the GLIMPSE survey. This diagram is used to
identify and classify objects with MIR colour excess (hereafter
denoted as CE). Within a region of $2\farcm5$ radius centred at
the NIR $H$-frame (see Fig. \ref{fig-cluster157Hdaofind}), a total
of 58 sources were simultaneously detected in the four GLIMPSE/IRAC-bands.
We identify 11 MIR sources with colour excesses, which are listed in 
Table \ref{MidIR YSOs}. Some of them are so obscured by dust that they 
have not been detected in the NIR bands.

Low-mass objects with MIR colour excess can be classified according 
to scheme by \cite{Allen et al 2004}. Class III objects, along with
back/foreground objects, have the MIR colours $[3.6] - [4.5] \approx 0$
and $[5.8] - [8.0] \approx 0$, Class II have colours within
$0 \la [3.6] - [4.5] \la 0.8$ and $0.4 \la [5.8] - [8.0] \la 1.1$,
whereas Class I objects are expected to present colours
$[3.6] - [4.5] \ga 0.7$ and $[5.8] - [8.0] \ga 1.0$. However, from the 
MIR photometry alone, we are not able to decide whether or not
these objects are low-mass YSOs.

Figure~\ref{YSO SEDs} presents the spectral energy distributions
(SEDs) of the MIR sources classified as YSOs, obtained by combining
their NIR and MIR magnitudes. Sources CE\,1 and CE\,10 present a high
spectral index $\alpha \equiv d\log{(\lambda F_{\lambda})}/d\log{\lambda}$
near $\lambda \sim 2\,\mu$m, which is characteristic of protostars in
an early stage of evolution \citep{Lada 1987}.

\subsection{The nature of IRAS\,16085-5138}
\label{The IRAS source}

IRAS\,16085-5138 is the only IRAS source in the field. According
to the classification scheme of \cite{Wood & Churchwell 1989}, its
IRAS colours are compatible with an ultra-compact H\,{\sc ii} region.
Methanol and hydroxyl masers have been detected at its vicinity
\citep[e.g.][]{Caswell et al 1980,Caswell et al 2000}. We identify
this source with CE\,10, which presents the highest spectral index
among all the CE-objects identified in this work, and is the
only one detected inside the IRAS position error ellipse.
Using photometric data from the 2MASS, GLIMPSE, and IRAS surveys, we
obtained the 2--100\,$\mu$m SED of this source (Fig.~\ref{IRAS SED}).
According to \cite{Lada 1987}, a spectral energy distribution rising
longward of $\lambda = 2\,\mu$m, such as that shown by this object,
is characteristic of Class I objects. We evaluated the spectral index
between $2$ and $25\,\mu$m (the dotted line in Fig.~\ref{IRAS SED}),
and found $\alpha \approx 3.6$, typical of YSO immersed in a dense
protostellar envelope, which is also in accordance with the classification
drawn from the MIR colour-colour diagram.

Since most of the energy of this object is emitted in the infrared,
the bolometric luminosity of the embedded protostar can be approximated
by integrating the SED between $2$ and $100\,\mu$m. Adopting the
distance of 3.29\,kpc and $A_V = 10$\,mag, we obtain a luminosity of $L=7.7\times10^3L_{\sun}$.
Since pre-main-sequence models \cite[e.g.][]{Iben 1965} indicate that
early-type stars evolve at constant luminosity from the Hayashi track
up to the main sequence, this result implies that the lower limit for
this stellar mass is $M \ga 10\,M_{\sun}$. This mass value corresponds
to a B0--B1 ZAMS star \citep[cf.][]{Hanson et al 1997}.

\subsection{MIR shell structure}
\label{MIR shell}

\begin{figure}
\includegraphics[width=0.48\textwidth,angle=0]{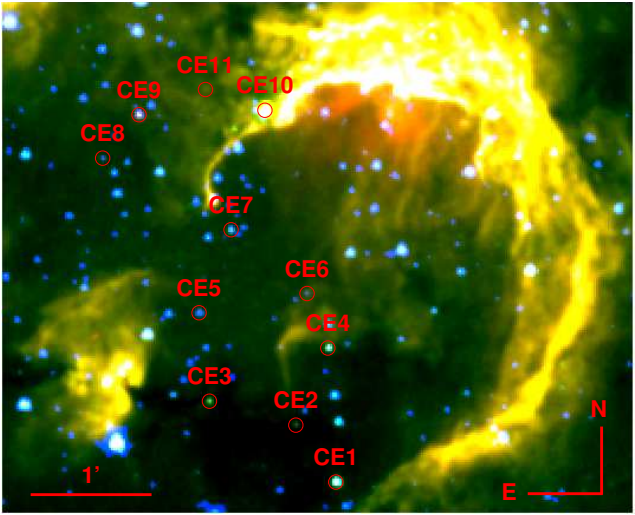}
\caption{RGB combination of $8.0\,\mu$m (R), $5.8\,\mu$m (G), and $3.6\,\mu$m (B) GLIMPSE images. The red circles indicate the objects showing MIR colour excess listed in Table \ref{MidIR YSOs}. In these three MIR bands, the peak emission coincides with the source CE\,10, which we identified as the GLIMPSE counterpart of IRAS\,16085-5138.}
  \label{RGB combination}
\end{figure}

Figure~\ref{RGB combination} presents a RGB combination of $8.0\,\mu$m (R), $5.8\,\mu$m (G), and $3.6\,\mu$m (B) GLIMPSE images that show a broken shell-like structure with a diameter of $\approx\!4\arcmin$ associated with GAL\,331.31-00.34. At the distance of 3.29\,kpc, this shell would have a diameter of $\approx\!3.8$\,parsec. This shell, identified as [CPA2006] S62 in the SIMBAD database, is named S\,62 in the catalogue by \cite{Churchwell et al 2006}, which contains over 300 MIR arcs. The authors interpreted the arcs as the projection on the sky of three-dimensional bubbles, and found that only one quarter of them would be associated with H\,{\sc ii} regions. In these cases, the bubbles would be expanding driven by the stellar wind and the radiation pressure from young massive stars. These stars would also power the bubbles, which are traced by the polycyclic aromatic hydrocarbon (PAH) emission produced in the photodissociation regions located at the edge of the H\,{\sc ii} regions. According to \cite{Deharveng et al 2010}, most of the bubbles are in fact associated with H\,{\sc ii} regions, and more than a quarter of them could be triggering the formation of a new generation of stars, as seems to be the case of RCW\,120 \citep{Deharveng et al 2009}.

In the case of GAL\,331.31-00.34, the IRAS source 16085-5138 and the methanol and hydroxyl masers \citep{Ellingsen et al 1996,Kuchar & Clark 1997, Walsh et al 1998,Caswell et al 2000,Caswell et al 1980}, typical tracers of ongoing massive star formation, are located at the bright northern border of the shell. At these positions, CS(2-1) and SiO molecular emission, which require densities higher than  $10^{4}$\,cm$^{-3}$, were detected by \cite{Bronfman et al 1996} and \cite{Harju et al 1998}, respectively, who carried out radio surveys towards IRAS sources with colours of UC-HII regions. This possibly indicates that the expansion of the northern part of the bubble has been hampered by a dense molecular environment. The same phenomenon may also occur in other parts of the bubble, but additional observations of CS and NH$_3$, which are good tracers of dense molecular regions, are still lacking.

We also find that the T Tauri and the Herbig Ae/Be candidates found from the NIR photometry are found preferentially towards the bubble, while the YSO candidates found from the MIR data are located in a limited north-south band crossing the openings of the broken shell.

\section{Summary}

We carried out $JHK$ photometry in the direction of the Galactic H\,{\sc ii}
region GAL\,331.31-00.34. We classified spectroscopically the main ionizing
sources and estimated their distance. As a result, we identified other
potential ionizing stars and performed their respective photometric
classifications. Data from the 2MASS, GLIMPSE, and IRAS surveys were
also explored. Our main findings are:
\begin{figure}
\includegraphics[width=0.48\textwidth,angle=0]{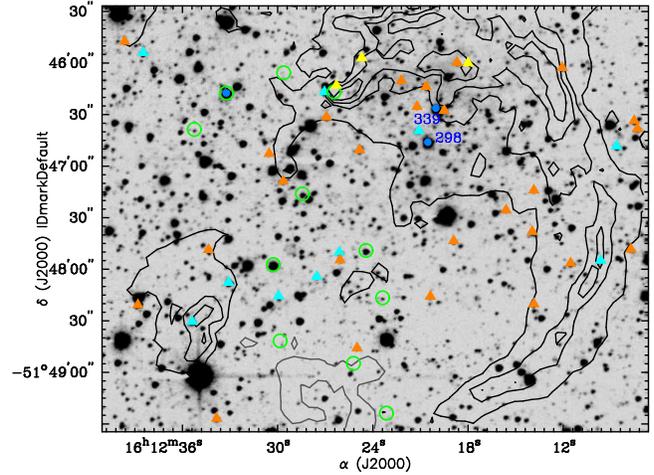}
\caption{Contour levels of the IRAC $8.0\,\mu$m image superimposed on $H$-band image. The yellow triangles show the positions of the methanol and hydroxyl maser sources. The cyan and orange triangles locate the low-mass and high-mass YSO candidates (possible reddened T-Tauri and the Herbig Ae/Be stars, respectively) classified according to the NIR photometry and listed in Table \ref{NIR YSOs}. The green circles indicate the objects showing MIR colour excess listed in Table \ref{MidIR YSOs}. The blue dots point to the hot stars spectroscopically identified.}
\label{8mu contours and star IDs}
\end{figure}
\begin{enumerate}

\item[i.] We identified, classified spectroscopically, and estimated
the spectroscopic parallax distances of two O-type stars associated with
this H{\sc ii} region: IRS\,298 (O6\,{\sc V}, $d = 3.35\pm0.61$\,kpc) and
IRS\,339 (O9\,{\sc V}, $d = 3.24\pm0.56$\,kpc).
Adopting the average distance of $3.29\pm0.58$\,kpc and comparing the
Lyman continuum luminosity of these stars with that required to ionize
the nebula, obtained from radio continuum observations, we concluded
that these two stars are the ionizing sources of GAL\,331.31-00.34.

\item[ii.] By analysing the NIR colour-colour diagram, 36 pre-main sequence (PMS) objects could be identified and classified: 9 T-Tauri and 27 Herbig Ae/Be candidates. From the GLIMPSE data, 11
objects with MIR colour excesses have been found.

\item[iii.] The MIR counterpart of the IRAS source 16085-5138 was
identified and its spectral energy distribution between $2$ and
$100\,\mu$m was analysed. We concluded that IRAS\,16085-5138 is
a massive YSO in an early stage of formation, 
with luminosity $L \ge 7.7\times10^3L_{\sun}$ and mass $M > 10\,M_{\sun}$.

\end{enumerate}

High spatial resolution observations in radio frequency of the ionized
and molecular gas are necessary for deeper studies of the dynamics and
evolution of star formation in the field of GAL\,331.31-00.34. Likewise, 
the possible connection between the diffuse high energy gamma-ray emission 
HESS J1614-518 and powerful winds and radiation fields of OB stars in the 
region should be subject of future hydrodynamic studies.

\section*{Acknowledgments}
This work was supported by the Brazilian agencies CAPES, CNPq and FAPESP. We wish to thank the staff of the Laborat\'{o}rio Nacional de Astrof\'{\i}sica for their assistance during the observations. ARL thanks the partial support by the ALMA-CONICYT Fund, under the project number 31060004, ``A New Astronomer for the Astrophysics Group -- Universidad de La Serena'', by the Physics Department, and by the Direcci\'on de Investigaci\'on Universidad de La Serena (DIULS), under program ``Proyecto Convenio de Desempe\~o CD11103''.

\newpage

\appendix

\section[]{}

\begin{figure*}
\centering\includegraphics[height=16cm,angle=-90]{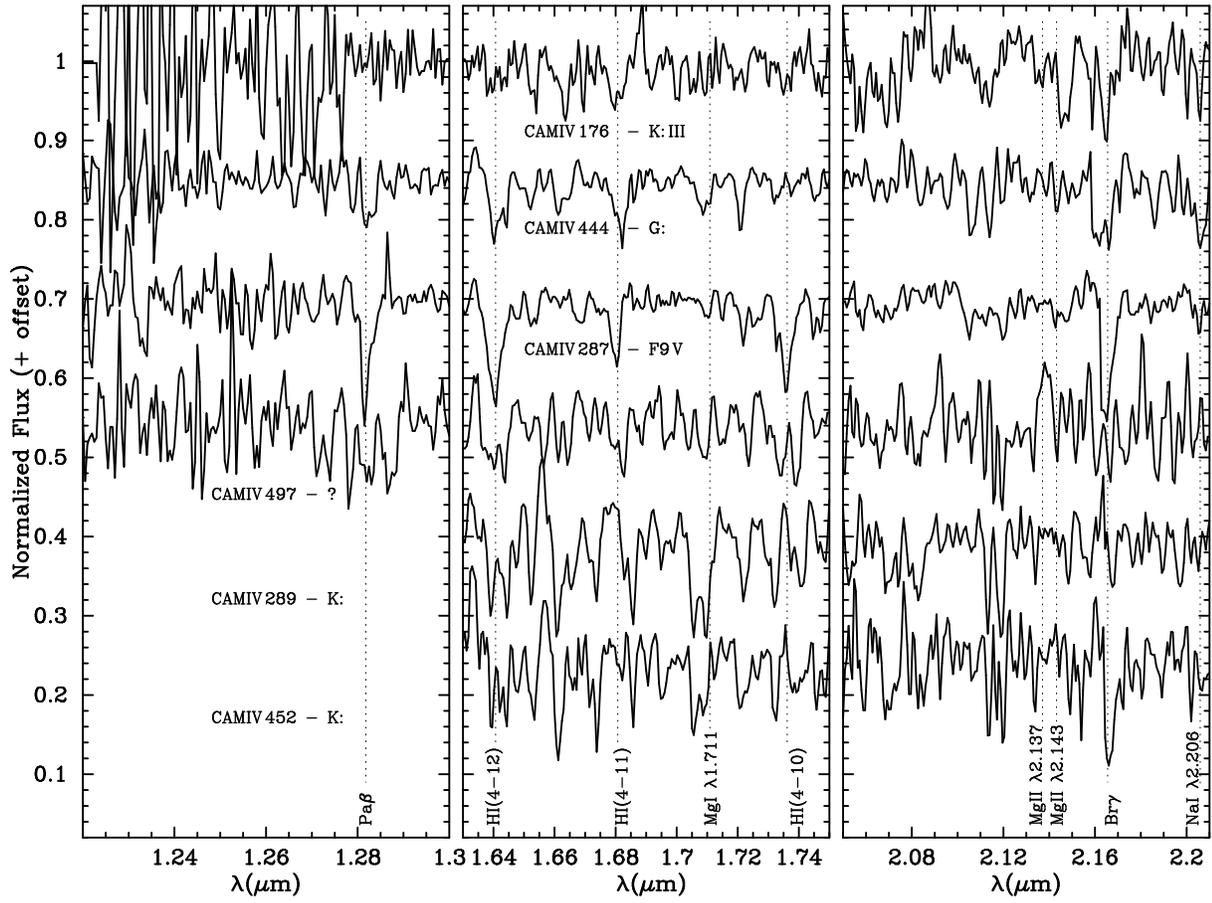}
\caption{Cold stars in the field of GAL\,331.31-00.34.}
\label{fig-spec-cold-stars}
\end{figure*}

\begin{table}
\caption{$JHK$ photometry}
\label{tab-photometry}
\tiny{
\begin{tabular}{@{}llllllllllll}
\hline\hline
IRS&\multicolumn{1}{c}{$\alpha$({\rm J2000})}&\multicolumn{1}{c}{$\delta$({\rm J2000})}&\multicolumn{1}{c}{$J$}&\multicolumn{1}{c}{$H$}&\multicolumn{1}{c}{$K_{\rm s}$}      \\
\hline
\noalign{\smallskip}
  3     &   16:12:33.028  & -51:49:23.11  &     16.166$\pm$0.070  &   13.485$\pm$0.018   &  12.246$\pm$0.066   \\
  41    &   16:12:26.166  & -51:49:27.27  &     13.479$\pm$0.028  &   11.951$\pm$0.016   &  11.199$\pm$0.037   \\
  42    &   16:12:33.848  & -51:49:26.83  &     13.792$\pm$0.031  &   13.405$\pm$0.014   &  13.046$\pm$0.112   \\
  79    &   16:12:14.127  & -51:49:03.79  &     11.658$\pm$0.029  &   9.809 $\pm$0.015   &  9.021 $\pm$0.015   \\
  87    &   16:12:37.336  & -51:48:57.01  &     13.042$\pm$0.025  &   11.841$\pm$0.012   &  11.447$\pm$0.039   \\
  94    &   16:12:24.999  & -51:48:45.83  &     14.965$\pm$0.031  &   14.212$\pm$0.014   &  13.090$\pm$0.231   \\
  98    &   16:12:39.840  & -51:48:41.18  &     9.475 $\pm$0.032  &   8.185 $\pm$0.013   &  7.682 $\pm$0.014   \\
  102   &   16:12:23.151  & -51:48:40.40  &     9.738 $\pm$0.028  &   9.044 $\pm$0.008   &  8.824 $\pm$0.015   \\
  108   &   16:12:34.752  & -51:48:33.93  &     15.580$\pm$0.039  &   13.894$\pm$0.017   &  12.969$\pm$0.085   \\
  113   &   16:12:31.255  & -51:48:30.62  &     12.930$\pm$0.028  &   11.270$\pm$0.009   &  10.601$\pm$0.018   \\
  114   &   16:12:35.383  & -51:48:30.43  &     15.455$\pm$0.038  &   13.179$\pm$0.018   &  11.801$\pm$0.046   \\
  117   &   16:12:28.621  & -51:48:29.84  &     12.435$\pm$0.029  &   11.468$\pm$0.014   &  11.163$\pm$0.022   \\
  133   &   16:12:13.855  & -51:48:20.25  &     14.057$\pm$0.028  &   13.593$\pm$0.022   &  13.071$\pm$0.112   \\
  135   &   16:12:38.748  & -51:48:16.20  &     10.923$\pm$0.029  &   10.217$\pm$0.013   &  9.954 $\pm$0.014   \\
  138   &   16:12:20.382  & -51:48:15.79  &     14.586$\pm$0.025  &   14.140$\pm$0.024   &  13.745$\pm$0.196   \\
  150   &   16:12:33.097  & -51:48:07.67  &     15.624$\pm$0.063  &   12.155$\pm$0.015   &  9.842 $\pm$0.013   \\
  164   &   16:12:13.044  & -51:47:59.20  &     13.887$\pm$0.029  &   12.279$\pm$0.016   &  11.643$\pm$0.028   \\
  168   &   16:12:30.291  & -51:47:57.57  &     11.302$\pm$0.026  &   10.720$\pm$0.015   &  10.448$\pm$0.013   \\
  169   &   16:12:31.902  & -51:47:57.47  &     10.657$\pm$0.028  &   9.965 $\pm$0.013   &  9.775 $\pm$0.016   \\
  172   &   16:12:09.642  & -51:47:54.88  &     15.896$\pm$0.060  &   14.828$\pm$0.042   &  13.973$\pm$0.360   \\
  176   &   16:12:24.451  & -51:47:49.26  &     14.321$\pm$0.026  &   12.807$\pm$0.015   &  12.347$\pm$0.081   \\
  178   &   16:12:34.366  & -51:47:48.61  &     13.576$\pm$0.029  &   13.295$\pm$0.010   &  12.752$\pm$0.131   \\
  184   &   16:12:18.908  & -51:47:43.46  &     14.291$\pm$0.026  &   13.787$\pm$0.014   &  13.393$\pm$0.197   \\
  196   &   16:12:40.637  & -51:47:37.13  &     11.914$\pm$0.032  &   10.663$\pm$0.012   &  10.148$\pm$0.015   \\
  197   &   16:12:13.960  & -51:47:38.10  &     13.890$\pm$0.030  &   13.448$\pm$0.024   &  12.800$\pm$0.116   \\
  206   &   16:12:36.513  & -51:47:32.56  &     12.382$\pm$0.027  &   11.490$\pm$0.016   &  11.162$\pm$0.021   \\
  217   &   16:12:19.360  & -51:47:28.72  &     9.105 $\pm$0.025  &   8.155 $\pm$0.011   &  7.927 $\pm$0.014   \\
  224   &   16:12:15.607  & -51:47:25.57  &     15.415$\pm$0.037  &   14.917$\pm$0.029   &  13.909$\pm$0.273   \\
  230   &   16:12:29.728  & -51:47:19.69  &     14.678$\pm$0.036  &   11.831$\pm$0.016   &  10.561$\pm$0.018   \\
  234   &   16:12:12.487  & -51:47:18.36  &     11.570$\pm$0.023  &   10.803$\pm$0.006   &  10.559$\pm$0.023   \\
  241   &   16:12:36.660  & -51:47:15.40  &     15.811$\pm$0.061  &   13.884$\pm$0.022   &  12.786$\pm$0.115   \\
  246   &   16:12:13.859  & -51:47:14.03  &     15.552$\pm$0.033  &   14.650$\pm$0.016   &  13.643$\pm$0.275   \\
  255   &   16:12:11.494  & -51:47:09.36  &     15.466$\pm$0.080  &   14.015$\pm$0.028   &  13.145$\pm$0.090   \\
  256   &   16:12:29.659  & -51:47:08.77  &     15.042$\pm$0.057  &   14.215$\pm$0.043   &  12.710$\pm$0.094   \\
  258   &   16:12:37.134  & -51:47:07.52  &     11.005$\pm$0.029  &   10.247$\pm$0.009   &  9.977 $\pm$0.019   \\
  262   &   16:12:32.924  & -51:47:05.09  &     12.240$\pm$0.026  &   11.318$\pm$0.010   &  10.978$\pm$0.021   \\
  264   &   16:12:28.519  & -51:47:04.38  &     11.391$\pm$0.026  &   11.089$\pm$0.012   &  11.095$\pm$0.025   \\
  268   &   16:12:30.324  & -51:47:01.98  &     11.528$\pm$0.024  &   10.457$\pm$0.014   &  10.035$\pm$0.016   \\
  277   &   16:12:26.742  & -51:46:57.38  &     11.536$\pm$0.026  &   11.224$\pm$0.016   &  11.200$\pm$0.031   \\
  278   &   16:12:34.600  & -51:46:56.74  &     10.497$\pm$0.027  &   9.805 $\pm$0.013   &  9.514 $\pm$0.017   \\
  283   &   16:12:30.538  & -51:46:52.68  &     16.162$\pm$0.085  &   15.411$\pm$0.050   &  14.346$\pm$0.311   \\
  285   &   16:12:27.736  & -51:46:50.82  &     15.592$\pm$0.056  &   13.071$\pm$0.010   &  12.046$\pm$0.043   \\
  286   &   16:12:19.625  & -51:46:51.10  &     15.052$\pm$0.025  &   12.684$\pm$0.015   &  11.686$\pm$0.041   \\
  287   &   16:12:29.116  & -51:46:50.29  &     11.203$\pm$0.025  &   10.169$\pm$0.009   &  9.736 $\pm$0.012   \\
  289   &   16:12:15.621  & -51:46:50.44  &     14.330$\pm$0.021  &   11.136$\pm$0.012   &  9.643 $\pm$0.013   \\
  295   &   16:12:32.209  & -51:46:46.38  &     15.284$\pm$0.037  &   11.904$\pm$0.011   &  10.302$\pm$0.013   \\
  298   &   16:12:20.539  & -51:46:46.07  &     11.674$\pm$0.024  &   10.288$\pm$0.008   &  9.593 $\pm$0.018   \\
  319   &   16:12:27.952  & -51:46:35.74  &     10.950$\pm$0.024  &   9.707 $\pm$0.012   &  9.216 $\pm$0.013   \\
  323   &   16:12:40.287  & -51:46:34.17  &     14.192$\pm$0.026  &   12.371$\pm$0.017   &  11.540$\pm$0.034   \\
  324   &   16:12:09.102  & -51:46:35.02  &     11.319$\pm$0.024  &   10.520$\pm$0.009   &  10.206$\pm$0.013   \\
  330   &   16:12:26.925  & -51:46:31.44  &     14.564$\pm$0.025  &   13.945$\pm$0.017   &  13.499$\pm$0.192   \\
  337   &   16:12:19.501  & -51:46:27.59  &     14.966$\pm$0.078  &   14.479$\pm$0.083   &  13.684$\pm$0.216   \\
  339   &   16:12:20.031  & -51:46:26.20  &     11.784$\pm$0.024  &   10.596$\pm$0.007   &  10.001$\pm$0.016   \\
  344   &   16:12:21.212  & -51:46:25.36  &     15.906$\pm$0.059  &   15.476$\pm$0.065   &  14.138$\pm$0.355   \\
  355   &   16:12:33.242  & -51:46:17.41  &     11.728$\pm$0.022  &   10.691$\pm$0.015   &  10.184$\pm$0.015   \\
  362   &   16:12:20.653  & -51:46:13.75  &     14.693$\pm$0.033  &   14.312$\pm$0.023   &  13.494$\pm$0.192   \\
  363   &   16:12:32.580  & -51:46:13.00  &     12.597$\pm$0.026  &   11.052$\pm$0.014   &  10.408$\pm$0.018   \\
  369   &   16:12:36.212  & -51:46:08.58  &     11.911$\pm$0.029  &   10.730$\pm$0.013   &  10.264$\pm$0.018   \\
  374   &   16:12:37.823  & -51:46:03.37  &     14.042$\pm$0.026  &   12.130$\pm$0.012   &  11.320$\pm$0.031   \\
  375   &   16:12:35.031  & -51:46:02.14  &     11.553$\pm$0.022  &   10.814$\pm$0.014   &  10.576$\pm$0.016   \\
  376   &   16:12:12.069  & -51:46:02.71  &     14.307$\pm$0.026  &   14.016$\pm$0.018   &  13.579$\pm$0.240   \\
  383   &   16:12:18.702  & -51:45:59.64  &     15.204$\pm$0.042  &   14.690$\pm$0.032   &  14.178$\pm$0.361   \\
  390   &   16:12:38.478  & -51:45:53.96  &     13.809$\pm$0.027  &   12.774$\pm$0.017   &  12.078$\pm$0.047   \\
  391   &   16:12:27.182  & -51:45:53.34  &     13.145$\pm$0.023  &   11.842$\pm$0.014   &  11.218$\pm$0.027   \\
  402   &   16:12:39.647  & -51:45:47.38  &     15.766$\pm$0.057  &   15.167$\pm$0.039   &  14.322$\pm$0.353   \\
  410   &   16:12:35.158  & -51:45:44.50  &     15.733$\pm$0.046  &   12.333$\pm$0.009   &  10.799$\pm$0.022   \\
  415   &   16:12:32.059  & -51:45:39.79  &     11.884$\pm$0.025  &   11.189$\pm$0.014   &  10.925$\pm$0.020   \\
  416   &   16:12:23.793  & -51:45:40.10  &     11.733$\pm$0.023  &   11.057$\pm$0.012   &  10.795$\pm$0.023   \\
  425   &   16:12:10.706  & -51:48:53.63  &     13.672$\pm$0.026  &   12.021$\pm$0.013   &  11.346$\pm$0.031   \\
  440   &   16:12:07.291  & -51:47:22.36  &     12.806$\pm$0.019  &   11.696$\pm$0.008   &  11.330$\pm$0.017   \\
  444   &   16:12:20.717  & -51:47:07.66  &     11.615$\pm$0.035  &   10.535$\pm$0.052   &  10.144$\pm$0.050   \\
  445   &   16:12:24.838  & -51:46:50.62  &     14.336$\pm$0.032  &   13.563$\pm$0.053   &  12.820$\pm$0.143   \\
  446   &   16:12:08.634  & -51:46:48.29  &     15.899$\pm$0.085  &   13.180$\pm$0.019   &  11.566$\pm$0.033   \\
  452   &   16:12:23.287  & -51:46:18.82  &     14.254$\pm$0.045  &   11.793$\pm$0.016   &  10.657$\pm$0.017   \\
  454   &   16:12:27.046  & -51:46:16.80  &     14.315$\pm$0.033  &   13.110$\pm$0.019   &  12.295$\pm$0.066   \\
  478   &   16:12:29.948  & -51:48:15.55  &     15.810$\pm$0.100  &   13.961$\pm$0.075   &  12.172$\pm$0.056   \\
  481   &   16:12:27.537  & -51:48:04.28  &     15.985$\pm$0.101  &   14.350$\pm$0.055   &  13.074$\pm$0.134   \\
  483   &   16:12:11.536  & -51:47:56.37  &     13.732$\pm$0.024  &   13.260$\pm$0.014   &  12.797$\pm$0.098   \\
  484   &   16:12:26.066  & -51:47:54.60  &     15.870$\pm$0.068  &   15.093$\pm$0.097   &  13.571$\pm$0.221   \\
  487   &   16:12:26.122  & -51:47:50.07  &     15.091$\pm$0.033  &   13.783$\pm$0.022   &  12.938$\pm$0.117   \\
  488   &   16:12:07.740  & -51:47:48.32  &     14.966$\pm$0.058  &   14.385$\pm$0.053   &  13.608$\pm$0.199   \\
  493   &   16:12:28.866  & -51:47:21.05  &     16.241$\pm$0.162  &   14.295$\pm$0.154   &  13.235$\pm$0.137   \\
  494   &   16:12:12.964  & -51:47:13.43  &     12.650$\pm$0.023  &   11.589$\pm$0.011   &  11.183$\pm$0.021   \\
  497   &   16:12:29.250  & -51:47:00.40  &     13.407$\pm$0.047  &   11.917$\pm$0.043   &  11.351$\pm$0.052   \\
  510   &   16:12:21.101  & -51:46:39.48  &     16.272$\pm$0.160  &   14.472$\pm$0.032   &  13.269$\pm$0.110   \\
  511   &   16:12:07.325  & -51:46:38.31  &     15.433$\pm$0.053  &   14.657$\pm$0.054   &  13.508$\pm$0.143   \\
  515   &   16:12:19.293  & -51:46:07.20  &     12.907$\pm$0.028  &   11.880$\pm$0.022   &  11.509$\pm$0.035   \\
  524   &   16:12:38.811  & -51:48:20.72  &     13.964$\pm$0.139  &   13.450$\pm$0.127   &  12.917$\pm$0.109   \\
  530   &   16:12:27.472  & -51:46:54.64  &     15.507$\pm$0.163  &   13.005$\pm$0.045   &  12.004$\pm$0.061   \\
  531   &   16:12:07.529  & -51:46:33.57  &     14.011$\pm$0.018  &   13.496$\pm$0.019   &  13.037$\pm$0.135   \\
  533   &   16:12:22.212  & -51:46:10.27  &     15.790$\pm$0.058  &   15.166$\pm$0.026   &  14.192$\pm$0.242   \\
\hline
\end{tabular}}
\end{table}

\end{document}